\documentclass[twocolumn]{aastex63}
\submitjournal{ApJS}

\usepackage{amsmath}
\usepackage{comment}

\newcommand{\RNum}[1]{\uppercase\expandafter{\romannumeral #1\relax}}

\graphicspath{{./}{figures/}}

\begin{document}



\title{The case of the missing VHE GRBs: \\
A retrospective study of \emph{Swift} gamma-ray bursts with Imaging Atmospheric Cherenkov Telescopes.}

\author{H. Ashkar}\thanks{corresponding author}
\affiliation{Laboratoire Leprince-Ringuet, École Polytechnique, CNRS, Institut Polytechnique de Paris, F-91128 Palaiseau, France}

\author{A. Sangaré}\thanks{corresponding author}
\affiliation{Laboratoire Leprince-Ringuet, École Polytechnique, CNRS, Institut Polytechnique de Paris, F-91128 Palaiseau, France}

\author{S. Fegan}\thanks{corresponding author}
\affiliation{Laboratoire Leprince-Ringuet, École Polytechnique, CNRS, Institut Polytechnique de Paris, F-91128 Palaiseau, France}

\author{J.~Damascene~Mbarubucyeye}
\affiliation{DESY, D-15738 Zeuthen, Germany}

\author{E.~Ruiz-Velasco}
\affiliation{Max-Planck-Institut f\"ur Kernphysik, P.O. Box 103980, D 69117 Heidelberg, Germany}

\author{S.J.~Zhu}
\affiliation{DESY, D-15738 Zeuthen, Germany}

\email{halim.ashkar@llr.in2p3.fr}
\email{aurelie.sangare@balliol.ox.ac.uk}
\email{sfegan@llr.in2p3.fr}


\begin{abstract}


Gamma-ray bursts (GRBs) are particle acceleration sites that can emit photons in the very high-energy (VHE) domain through non-thermal processes.  From 2004 until 2018, the current generation of Imaging Atmospheric Cherenkov Telescopes (IACTs) did not detect any GRB in the VHE domain. However, from 2018 to 2020, five detections have been reported. In this work, we try to solve the case of the missing VHE GBRs prior to 2018. We aim to identify GRBs that might have eluded VHE detection in the past years by the H.E.S.S., MAGIC, and VERITAS IACTs. To do so, we study GRBs with known redshift detected by \emph{Swift} from 2004 until June 2022. We first identify all GRBs that could have been observed by these IACTs since 2004, considering observation conditions and visibility constraints. We assume a relation between the X-rays and the VHE gamma rays based on the VHE GRBs detected to date and combine this with the redshift measurements, instrument response information, and observation conditions to predict the observed VHE gamma-ray flux from the \emph{Swift}-XRT measurements. We report findings on 12 bright low-redshift GRBs that we identify as most likely to have been detected in the VHE domain by current IACTs. The rate of IACT-detectable GRBs with ideal observation conditions is $<$1 VHE GRB per year with the current configuration. With its lower energy threshold and higher sensitivity, this rate increases to $\sim$4 VHE GRBs per year with CTA.   \\

\end{abstract}


\section{Introduction}
\label{sec:introduction}
Gamma-ray bursts (GRBs) are bright flashes of electromagnetic radiation from astrophysical origin. There are two categories of GRBs, long GRBs, which can be produced by a subset of core collapse of massive stars, and short GRBs which can be produced by the merger of neutron stars. GRBs are sites for particle acceleration and intense magnetic field where non-thermal emission is produced via  the synchrotron and inverse-Compton processes. These non-thermal processes can produce gamma rays that can potentially reach the very high energy (VHE) regime ($\geq 100$ GeV). A few photons from GRBs such as GRB130427A~\citep{Ackermann_2014} have been detected at GeV energies by the Large Area Telescope (LAT) on board the \emph{Fermi} observatory. Although current-generation Imaging Atmospheric Cherenkov Telescopes (IACTs) have been observing GRBs in the VHE regime (VHE GRBs in the following) for almost two decades  - i.e. since the installation of the first telescope of the High Energy Stereoscopic System (H.E.S.S.) in 2002 -,  no significant detection above 200 GeV was made for 17 years. 

In 2019, the detections of two VHE GRB afterglows were published: GRB\,180720B~ by the~\cite{Abdalla_2019} and GRB\,190114C by the~\cite{2019Natur.575..455M, 2019Natur.575..459M}. These were followed in 2021 by the H.E.S.S. VHE GRB\,190829A~\citep{HESS2021} detection. The three GRBs showed in the afterglow phase a comparable energy flux between the X-rays in the 0.3-10 keV range and the VHE gamma rays in the TeV range, as well as a similar decaying behavior over time in these energy bands. They also showed that the photon index in the VHE domain is approximately $\mathrm{\gamma = 2}$ for all of them. These three detections in a span of $\sim$1 year, when compared to the 16 years of non-detection, prompt inquiry. Why were no VHE GRBs detected before 2018~\citep{2019GCN.23701....1M, 2019ATel12390....1M}? Did observation strategies and instrument upgrades play a role in the recent detections of VHE GRBs? How many potential VHE GRBs were missed in the past? Are VHE GRBs particularly rare? 

In this work, we retrospectively inspect all GRBs detected by the Neil Gehrels \emph{Swift} Observatory. These GRBs have a localization uncertainty of a few arcminutes to a few arcseconds. In comparison, IACTs have a field of view that spans several square degrees in the sky, making the coverage of the GRB regions relatively easy. We consider three currently active IACTs: H.E.S.S., MAGIC, and VERITAS. We look back at all the \emph{Swift} GRB alerts that these observatories could have followed using current observation criteria and observation delays. The aim is to find those that could have been detected in the VHE domain based on the GRB X-ray flux, GRB distance, observation conditions, and telescope sensitivity, regardless of what observations were made.

In Sec.~\ref{sec:retro_vis}, we inject GRB alerts into a pipeline~\citep{hessTOOsystem} that takes into consideration telescope observation and visibility conditions to identify the GRBs that could have been observed by the three IACTs. In Sec.~\ref{sec:assumptions}, we present the hypotheses that we exploit on the GRBs' X-ray and gamma-ray emission, together with the underlying methodology that we follow. Sec.~\ref{sec:results} presents the results of the analysis, discusses caveats, and identifies potential GRBs of interest. In Sec.~\ref{sec:discussion}, we discuss the results and answer the questions presented in this section. We also extend our study to future IACTs, before concluding in Sec.~\ref{sec:conclusion}.

\section{Retrospective observation simulations}
\label{sec:retro_vis}
\begin{figure*}[!htb]
  \centering
\includegraphics[width=1\textwidth]{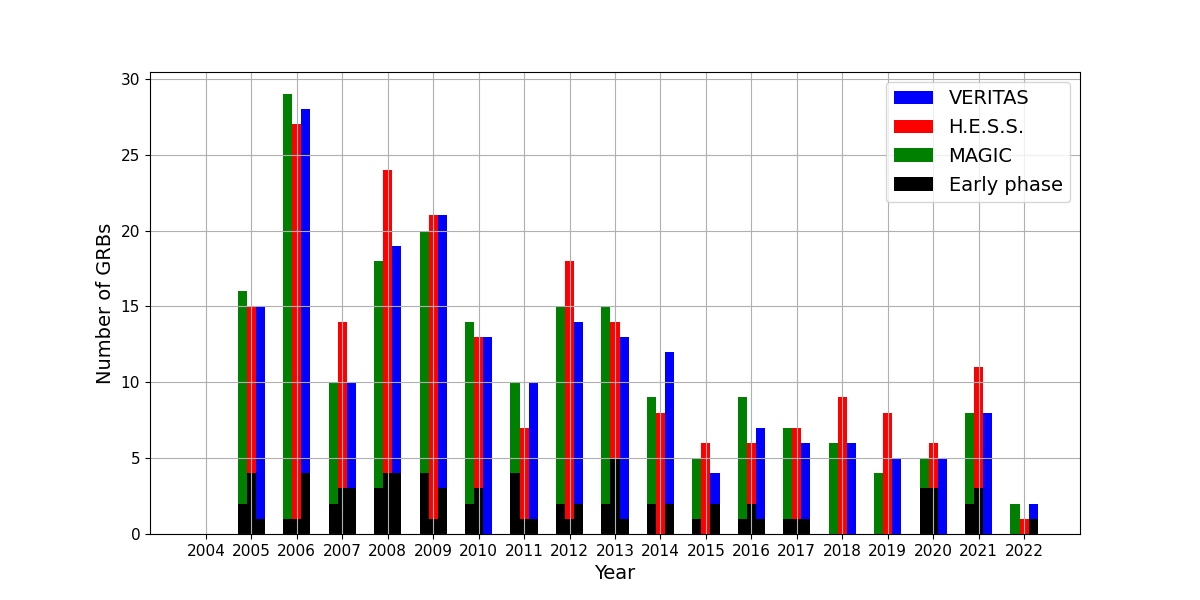}
\caption{Number of potential. GRB observation per year for H.E.S.S., MAGIC, and VERITAS. The GRBs that could have been observed with a delay of less than 600 seconds are marked in black as Early phase observations. We note that the MAGIC and VERITAS configurations considered in this work are only valid after 2007 and 2009 respectively and that H.E.S.S. only started implementing moonlight observations in 2019.}
\label{fig:GRB_OBS}
\end{figure*}

GRB alerts are distributed via networks and brokers like the GCN network~\footnote{\url{https://gcn.gsfc.nasa.gov/swift_grbs.html}}. To reproduce the reception of GRB alerts, we first retrieve all available \emph{Swift} GRB notices since 2004 from the GCN network. In 2014, the alert formats changed to VoEvent2.0~\citep{allan2017voevent}. These alerts are then injected in a pipeline simulating a telescope's respective observation and visibility conditions at the time of arrival of the alerts. The pipeline filters out alerts that do not match telescope observation criteria and keeps the ones that do. Given that collaborations have different observation condition requirements that can change with time, and given that these preferences information  are not always public, we choose to standardize observation criteria for all three IACTs in this work. These criteria are summarized in Tab.\ref{tab:obs_cond}. IACTs, are very sensitive to the night sky background light and generally, operate under total darkness, in the absence of the moon. However, to increase the duty cycle of the telescope, operations under moderate moonlight brightness were implemented by the H.E.S.S., MAGIC, and VERITAS collaborations. The MAGIC cameras were designed from the beginning (2007) to be able to operate under moderate moonlight conditions, when the phase of the Moon does not exceed 50\%~\citep{2017APh....94...29A}. Above this value, the camera would have to be operated in a reduced high voltage (HV) mode due to the increased night sky brightness. The VERITAS telescopes can operate under nominal conditions for Moon illumination within 35\%. VERITAS implemented in 2012 a reduced HV mode that allows the telescope to observe under up to 65\% Moon illumination~\citep{2015ICRC...34..989G}. Both MAGIC and VERITAS can use UV filters that allow them to operate under bright moonlight up to 80\% illumination, with a significant cost on the sensitivity. H.E.S.S. on the other hand did not implement moonlight observations until 2019~\citep{2023NIMPA105568442O}. The conditions outlined in Tab.~\ref{tab:obs_cond} reflect the current status of the H.E.S.S. observation conditions. Since we are looking for missed opportunities, in the following we will consider that these IACTs can observe under moonlight since the beginning of their operations. We will use conservative observation conditions to maintain the sensitivity of the telescopes. The alerts are injected into the pipeline anew for each different telescope. The three IACTs, located at three different latitudes, provide coverage of both hemispheres. From 2004 until May 2023, we find 1008, 913, and 886 notices that fulfill the observation criteria in Tab.~\ref{tab:obs_cond} for H.E.S.S., MAGIC, and VERITAS respectively. For those, the trigger number, event time, coordinates and observations windows are saved. To filter out fake alerts, such as noise and flares from known sources, the trigger should be flagged as a new GRB, a point source-like, not found in any catalog, and should be identified as a GRB on the ground. Amongst the remaining alerts, we keep only the GRBs reported in the \emph{Swift} GRB catalog~\citep{Lien_2016}. Until mid-July 2022, 1527 GRBs are included in this catalog.  We further restrict our analysis to GRBs that have redshift measurement\footnote{\url{https://swift.gsfc.nasa.gov/results/batgrbcat/index_tables.html}}. A total of 488 GRB redshifts are reported until June 2022 (with some uncertain values that we discard for the following). GRBs with at least two successive points of \emph{Swift}-XRT X-ray data are considered. The number of potential observations per year is shown in~\ref{fig:GRB_OBS}. From 2004 until June 2022, 215, 201, and 198 GRBs are kept for H.E.S.S., MAGIC, and VERITAS, respectively (noting that GRBs may be observable by more than one IACT). We choose to use data for the full observation window, although some collaborations might choose not to spend the entire available observation time during a night on GRB observations. 

\begin{table}
\small
\begin{tabular}{ll}
  \hline
  
    Maximum allowed zenith angle   & 60 deg \\  
    Field of view   & 2 deg \\  
    Maximum allowed observation delay   & 24 hours \\
    Maximum Sun altitude & -16 deg \\
    Maximum Moon Phase &  40\% \\ 
    Maximum Moon altitude & 65 deg \\
    Minimum Moon-source separation  & 45 deg\\
    Maximum Moon-source separation & 145 deg\\
    Minimum observation duration & 6 minutes\\
   \hline
   \hline
\end{tabular}
\caption{Observation and visibility conditions used to select observable alerts at the time of reception by H.E.S.S., MAGIC, and VERITAS. In order to be considered observable, an alert follow-up observation must satisfy all the above conditions.}
\label{tab:obs_cond}
\end{table}

\section{Assumptions and methodology}
\label{sec:assumptions}
\subsection{Intrinsic VHE gamma-ray flux}
For this study, we make three assumptions about the VHE emission, motivated by the three aforementioned detected VHE GRBs. The first assumption is that the unabsorbed X-ray and VHE gamma-ray energy fluxes are related by $\mathrm{\phi_{\gamma}^{(u)} \times F \equiv  \phi_{X}^{(u)} }$, where $\mathrm{F}$ is a factor deduced from observations of the three detected VHE GRBs. The former can vary between 1 and 3. For example, in the case of GRB\,180720B, at around 10 hours after the burst, $\mathrm{F \sim 1}$ up to 440 GeV. For GRB\,190114C, in the early afterglow phase, $\mathrm{F \sim 1.5}$ and later on increases to $\mathrm{F \sim 2.5}$ up to 1 TeV. For GRB\,190829A, at around 4 hours after the burst, $\mathrm{F \sim 3}$ between 0.2 and 4 TeV ($\mathrm{E_1^{(u)}}$ and $\mathrm{E_2^{(u)}}$ respectively). To start with, we take the unabsorbed X-ray flux between 0.3-10 keV to be three times higher than the intrinsic VHE flux between 0.2-4 TeV band (i.e. $\mathrm{F=3}$), as was the case for GRB\,190829A. The latter GRB was considered because it is the closest GRB detected at VHEs. Consequently, among the three detected VHE GRBs, it is the least affected by gamma-ray absorption by the extragalactic background light (EBL). The second assumption is that the temporal behavior of the X-rays and VHE gamma rays follows a power law decay and both decays are similar. The decay indices in the afterglow phase are similar, giving $\mathrm{\alpha_{\gamma} =  \alpha_{X}}$. To get the X-ray fluxes, we use the \texttt{swifttools}\footnote{\url{https://www.swift.ac.uk/API}} package to query the \emph{Swift} Burst Analyser data. For details of how these light curves were produced, see~\cite{Evans_2010}. The X-ray light curves are then fitted using a decaying power law function following $\mathrm{\phi_{X}^{(u)} \propto t^{\alpha_{X}}}$. We only fit the X-ray data starting 2000 seconds after the GRB alert was triggered. This is motivated by the fact that we found that this is a reliable marker to characterize the onset of X-ray afterglow emission decay as a power law function. We note, that the choice of the 2000 seconds marker is a convention that we adopt for this work and is not based on the start of the afterglow phase.  The third assumption is that the photon index in the VHE band is $\mathrm{\gamma = 2}$. In what follows, the observed VHE gamma-ray flux is calculated from the intrinsic energy flux, taking into consideration EBL absorption effects, IACT effective areas, and background rates.  

\subsection{EBL absorption}
The low energy photon fields, including the EBL absorption of gamma rays via electron-positron pair production, result in a gamma-ray rate observed on Earth lower than that expected from a $\mathrm{1/D^2}$ level, where $\mathrm{D}$ is the distance of the source. We note that other factors such as the angular opening of a GRB jet and the possible tilt angle towards the observer also influence the observed gamma-ray rate. The gamma-ray absorption increases  with energy and redshift and becomes dominant around a redshift of $\mathrm{z=0.3}$. EBL absorption plays a major role in the lack of detection of VHE gamma rays from GRBs since most GRBs are distant, with a GRB redshift distribution for \emph{Swift} peaking above $\mathrm{z=1}$. The redshift dependency of EBL absorption justifies restraining the study to GRBs with known redshift only. We use the Dominguez EBL absorption model~\citep{2011MNRAS.410.2556D}. The VHE gamma-ray flux on Earth becomes: $\mathrm{\phi_{\gamma}^{(e)} =  \phi_{\gamma}^{(u)} e^{-\tau(E, z)}}$, where $\mathrm{e^{-\tau(E, z)}}$ is the energy- and redshift-dependent EBL absorption coefficient.

\subsection{Effective area}
The effective area can vary largely with the zenith angle of a GRB. We search for public information on the effective areas for H.E.S.S., MAGIC, and VERITAS. For each IACT we consider two epochs. For H.E.S.S., the first epoch is from 2004 to 2012 - when H.E.S.S. was composed of four 12-m telescopes (H.E.S.S. \RNum{1}) - and the second one is from 2012 to 2022 - after a fifth 28-m telescope was added to the array (H.E.S.S. \RNum{2}). We also consider the 28-m alone separately (H.E.S.S. MONO) in some cases. From~\cite{Aharonian_2006}, we find three effective areas for the H.E.S.S. \RNum{1} telescopes for zenith angles of 20, 45, and 60 degrees. For H.E.S.S. \RNum{2}, we take the effective area of all five telescopes working together in the \texttt{Combined Analysis} configuration computed from simulations at around 20 degrees zenith angle~\citep{holler2015photon}. We use the scaling between the three effective areas of H.E.S.S. \RNum{1} in order to compute new effective areas for H.E.S.S. \RNum{2} for 45 and 60-degree zenith angles. For MAGIC, the first epoch is from 2009 to 2012 - after the two 17-m MAGIC telescopes were installed (here MAGIC \RNum{1}) - and the second one is from 2012 to 2022 - after they underwent major upgrades (here MAGIC \RNum{2}). The effective area for the first epoch for zenith angles between 0 and 30 degrees and the second epoch for zenith angles between 0 and 30 and 30 and 45 degrees are taken from~\cite{Aleksi__2016}. The scaling from MAGIC \RNum{2} is used for MAGIC \RNum{1} to compute effective areas for 30 and 45-degree zenith angles. To get the effective area between 45 and 60 degrees for MAGIC \RNum{1} and \RNum{2}, we use the H.E.S.S. effective area scaling from 45 to 60 degrees zenith angles. We use the scalings from MAGIC for H.E.S.S. MONO (they are both large collection area telescopes). For VERITAS, the first epoch is from 2007 to 2012, and the second one is from 2012 to 2022 (here VERITAS \RNum{1} and \RNum{2}). There is one available effective area at 20 degrees zenith angle for each epoch\footnote{\url{https://veritas.sao.arizona.edu}}. Since no more than one effective area is available for VERITAS, we use the scaling used for the H.E.S.S  \RNum{1} telescopes (they have a similar four-telescopes configuration). For each telescope and each epoch, we use three effective areas at different zenith angles.

\subsection{Energy bounds and background rates}
The instrument energy range is defined between $\mathrm{E_1}$ and $\mathrm{E_2}$, which differ from the intrinsic energies $\mathrm{E_1}^{(u)}$ and $\mathrm{E_2}^{(u)}$ used to deduce the VHE energy flux from the X-ray flux. The flux $\mathrm{\phi_{\gamma}^{(e)}}$ reaching the IACTs is integrated between $\mathrm{E_1}$ and $\mathrm{E_2}$ to determine the observed flux. Due to the large collecting area of the H.E.S.S. 28-m and the MAGIC 17-m telescopes, we use a lower energy threshold. Tab.~\ref{tab:available_info} shows the different energy ranges used for each IACT, alongside other analysis parameters.   

\begin{table*}
\centering
\small
\begin{tabular}{ccccccc}
  \hline
  IACT & $\mathrm{E_1}$ (TeV) & $\mathrm{E_2}$ (TeV) & Reference for eff. area & $\mathrm{\alpha}$ & Bkg. rate (Hz)  & Zenith (deg) \\
    \hline
  H.E.S.S. \RNum{1} & 0.2 & 4 & \cite{Aharonian_2006}& 0.14 & 0.0865  & 45\\
  H.E.S.S. \RNum{2} & 0.1 & 4 & \cite{holler2015photon}  & 0.08278 & 0.1287  & 45\\
  H.E.S.S. MONO & 0.1 & 4 & \cite{holler2015photon} & 0.102 & 0.0837  & 45\\
  VERITAS \RNum{1} & 0.2 & 4 & \url{https://veritas.sao.arizona.edu} & 0.14 & 0.07951  & 20\\
  VERITAS \RNum{2} & 0.2 & 4 & \url{https://veritas.sao.arizona.edu} & 0.14 & 0.1101  & 20\\
  MAGIC \RNum{1} & 0.1 & 4 & \cite{Aleksi__2016} & 0.2 & 0.49  & $<$ 30 \\
  MAGIC \RNum{2} & 0.1 & 4 & \cite{Aleksi__2016}  & 0.2 & 0.41  & $<$ 30\\
   \hline
   \hline
\end{tabular}
\caption{Analysis parameters for computing the observed VHE gamma-ray flux by IACTs. For each IACT, the energy bounds, $\mathrm{E_1}$ and $\mathrm{E_2}$, the reference to existing effective areas, the ratio $\mathrm{\alpha}$ and the background rate at a given zenith angle are displayed.}
\label{tab:available_info}
\end{table*}

In VHE gamma-ray astronomy, cosmic rays constitute a substantial background. Most analysis methods use filtering methods that discriminate most background events based on the shower image parameters seen by the IACT cameras (e.g.~\cite{1985ICRC....3..445H}). Background estimation methods are needed to identify the degree of contamination of the VHE gamma-ray signal~\citep{Berge_2007}. ON and OFF regions are defined as the source and background regions respectively. The number of events in each region is $\mathrm{N_{ON}}$ and $\mathrm{N_{OFF}}$ respectively. The size (and acceptance) ratio between the ON and OFF regions is defined as $\mathrm{\alpha}$. The background rate (after filtering) for each telescope is taken from available public information. For H.E.S.S. \RNum{1}, we consider a background rate at low zenith angles of $\mathrm{R^B = 0.08658}$ events/second following the standard \texttt{Ring Analysis}~\citep{Aharonian_2006} with $\mathrm{\alpha} = 0.14$. These are deduced from data taken on the Crab Nebula at high zenith angles. For H.E.S.S. \RNum{2}, we compute the background rate from the Crab observations in~\cite{holler2015photon} with $\mathrm{R^B = \frac{N_{OFF}}{Live~time} =  0.213}$ and $\mathrm{\alpha} = 0.08278$. We also use $\mathrm{R^B =  0.0837}$ events/second and $\mathrm{\alpha} = 0.102$ for H.E.S.S. MONO. After filtering, background events mainly from protons are dominant, followed by other nuclei and electrons (and positrons). The spectral distribution for protons and electrons can be approximated by a $\mathrm{E^{-2.7}}$  and $\mathrm{E^{-3.3}}$ power law respectively. In order to get the background rate at different zenith angles, we integrate an $\mathrm{E^{-3.3}}$ cosmic-ray spectrum between $\mathrm{E_1}$ and $\mathrm{E_2}$. This is a conservative approach that overestimates the background at low energies, where most of the signal is expected. We multiply it by the effective area available at the given zenith angle for each telescope and each epoch and use the results to scale the background rate at different zenith angles. Considering a $\mathrm{E^{-2.7}}$ background spectrum does not heavily impact the results because the background rates decrease by around a quarter for high zenith angles and increase by a factor of $\sim$2 for high zenith angles. Since VERITAS is an array of four 12-m telescopes similar to H.E.S.S., we use the H.E.S.S. \RNum{1} background rates for VERITAS. To scale the background rate to the VERITAS level, we use the  scaling (assuming an $\mathrm{E^{-3.3}}$ cosmic-ray spectrum)  between the effective areas of the two IACTs, at 20 degrees zenith angles this time. For MAGIC \RNum{1} and \RNum{2}, the integral background rates at different energy thresholds and zenith angle observations are available from~\cite{Aleksi__2016} respectively. For the missing information, like for the other telescopes, the background rates are computed by scaling the cosmic-ray spectrum to the respective effective areas. 

\subsection{Observed VHE gamma-ray energy flux}
The relation between the unabsorbed X-ray and VHE gamma-ray energy fluxes at a given time is : 
\begin{equation}
\label{eq:equality}
\mathrm{\phi_{\gamma}^{(u)} \times F=  \int_{E_1^{(u)}}^{E_2^{(u)}} E \frac{dN}{dE} dE \equiv \phi_{X}^{(u)}}
\end{equation}
where $\mathrm{E_1^{(u)}}$ and $\mathrm{E_2^{(u)}}$ are the energy bounds (0.2 and 4 TeV respectively) and  $\mathrm{\frac{dN}{dE}}$ is the photon spectrum, with: 
\begin{equation}
  \mathrm{  \frac{dN}{dE} =  N_0 \left(\frac{E}{E_0}\right)^{-\gamma}}
\end{equation}
$\mathrm{N_0}$ is the normalization constant to be found using equation \ref{eq:equality}. $\mathrm{E}$ is the energy and $\mathrm{E_0}$ is the reference energy, taken as 1 TeV. $\mathrm{\gamma}$ is the photon index with $\mathrm{\gamma} = 2$.  
The signal rate of the observed VHE gamma rays is given by: 

\begin{equation}
 \mathrm{   R(t) = R_0 \hspace{0.05cm} t ^{\alpha_{X}}}
\end{equation}
with
\begin{equation}
 \mathrm{   R_0 = \int_{E_1}^{E_2} \frac{dN}{dE}(t) e^{-\tau_{EBL}} A(E) dE}
\end{equation}
where $\mathrm{A(E)}$ and $\mathrm{e^{-\tau_{EBL}}}$ are the effective area and the EBL absorption coefficient respectively, both dependent on energy. $\mathrm{E_1}$ and $\mathrm{E_2}$ are the energy bounds considered for each instrument. The observed VHE gamma-ray signal becomes 
\begin{equation}
 \mathrm{   S = \int_{t_1}^{t_2} R(t) dt}
\end{equation}

\begin{figure}
  \centering
\includegraphics[width=0.45\textwidth]{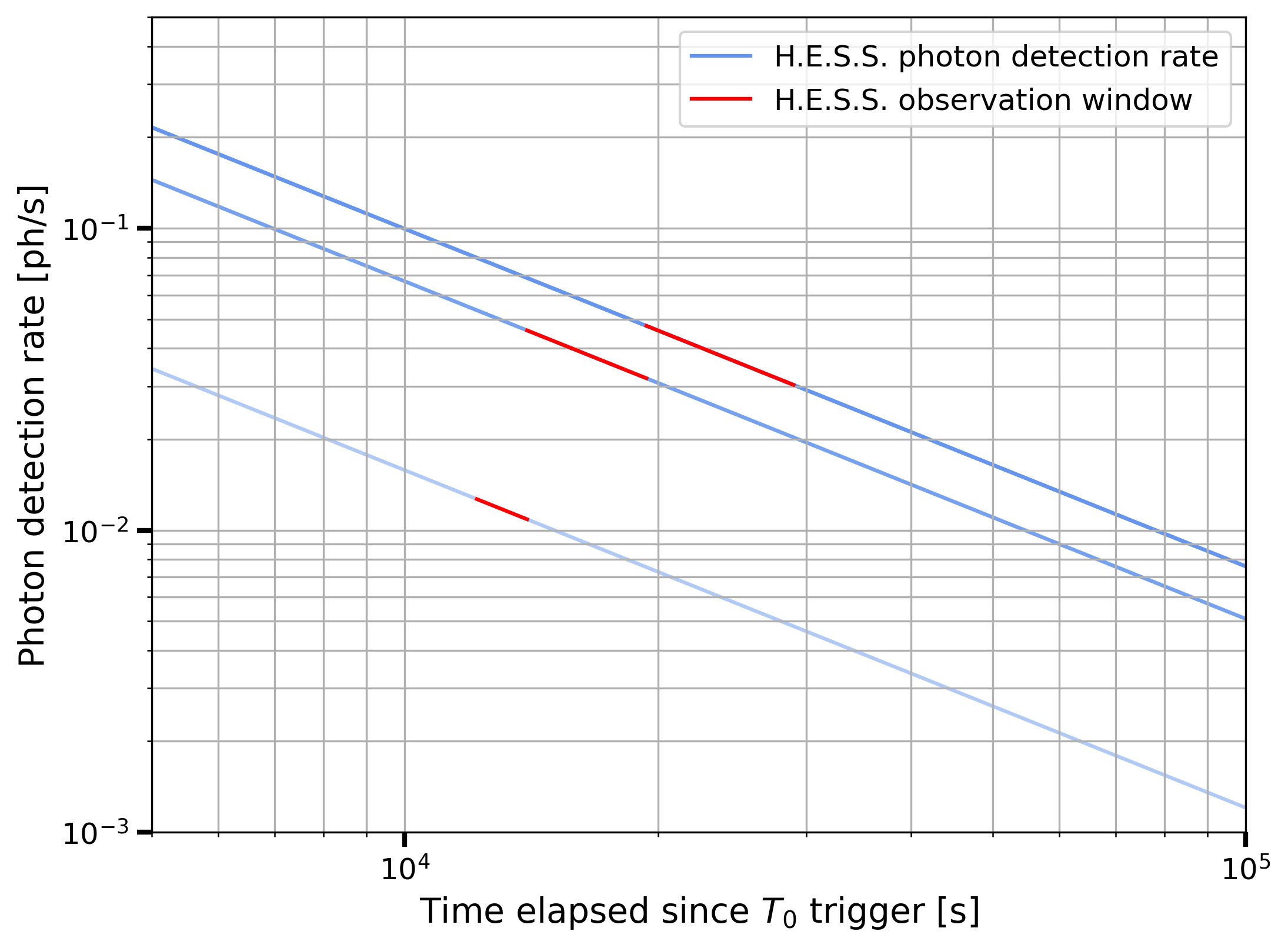}
\caption{Photon rates of GRB\,190829A as a function of time. The blue lines correspond to the photon rates starting at the first available X-ray data from \emph{Swift}-XRT. The red lines cover the H.E.S.S. simulated observation windows. Three photon rates are shown as three different zenith angle-dependent effective areas are considered as the source changes positions in the sky.}
\label{fig:HESS_rates}
\end{figure}

The background rate is: 

\begin{equation}
  \mathrm{  B = \int_{t_1}^{t_2} R^B dt}
\end{equation}

where $\mathrm{t_1}$ and $\mathrm{t_2}$ are the observation's start and end times respectively. Since the effective area and background rate are dependent on the zenith angle, we split the time of observation into steps of 30 minutes for each simulated observation. Every 30 minutes, the zenith angle of the simulated observation is recomputed, and the effective area and background rate for the angle matching the closest respective zenith angle at the beginning of the window are used.  We calculate $\mathrm{S}$ and $\mathrm{B}$ for each time window and then sum them. In Fig.~\ref{fig:HESS_rates}, we present the signal rates expected from GRB\,190829A that would have been observable by H.E.S.S. in blue and the simulated H.E.S.S. observation windows in red. In the case of GRB\,190829A, the observations start at a zenith angle of 60 degrees. The source then climbs in the sky to an apporximate 15-degree zenith angle. With three available effective areas at 20, 45, and 60 degrees zenith angle, three different photon rate curves are displayed (in blue), each corresponding to a different value of $\mathrm{N_0}$ and a different zenith angle. In each time window, the signal $\mathrm{S}$ is calculated by integrating between $\mathrm{t_1}$ and $\mathrm{t_2}$. The last observation window is shorter than 30 minutes and corresponds to the time remaining after the full 30-minute windows are computed. Note that in some cases, the observations start before any X-ray data is available.

The average number of events in the ON region is $\mathrm{N_{ON} = S + \alpha B}$. The average number of events in the OFF region is $\mathrm{N_{OFF} = B}$. We  calculate the significance of detecting a VHE gamma-ray signal using the formalism presented by~\cite{1983ApJ...272..317L}.

\section{Results}
\label{sec:results}
We test our method on GRB\,190829A with the H.E.S.S. \RNum{1} configuration. In reality, the H.E.S.S. observations started 4.3 hours after the burst and lasted for a total of 3.6 hours during the first night of observations. The significance of detection for the first night is $\mathrm{21.7 \sigma}$. In our simulations, the observation window is larger, since observations started earlier. We adjust our observation window to the one reported by the~\cite{HESS2021}. We find a significance of detection of $\mathrm{20.9 \sigma}$ for the first night with $\mathrm{N_{ON} =682}$. Doubling the background rate lowers the significance to $\mathrm{16.5 \sigma}$. For the case of GRB\,180720B, we consider $\mathrm{F = 1}$, as seen in~\cite{Abdalla_2019}.  We also reduce the observation window to match the window reported in the same publication. We get a $\mathrm{2.6 \sigma}$ significance, assuming a H.E.S.S. analysis with the 28-m telescope only, whereas a $\mathrm{5.3 \sigma}$ was reported by H.E.S.S. using the same analysis but with \texttt{loose cuts}. Extending to the full observation window, we get $\mathrm{3.7 \sigma}$. In the following, the detection threshold is set at $\mathrm{2 \sigma}$. 

The difference between the results reported here and the results reported in the detection papers by H.E.S.S. may result from several systematics. The first to be considered is the factor $\mathrm{F}$ considered here. We also do not differentiate between short, long, or ultra-long GRBs. Some GRBs, like the ultra-long GRB\,130925A~\citep{Evans_2014}, require special treatment to ensure that the X-ray light curve points do not exceed our fit in the observation window. The second caveat to be considered is the assumption on the photon index $\mathrm{\gamma}$. We have considered $\mathrm{\gamma} = 2$. Taking $\mathrm{\gamma} = 2.5$ for the test case of GRB\,190829A described above reduces the significance to $\mathrm{18.4 \sigma}$ with $\mathrm{N_{ON} =621}$. Doubling the background rate for the $\mathrm{\gamma} = 2.5$ case reduces the significance to $\mathrm{14.2 \sigma}$. Moreover, we do not account for dead-time (camera, slewing) instrumental failures or weather issues which reduce the live time for the calculation of $\mathrm{S}$ and $\mathrm{B}$, and hence reduce these two values simultaneously. We also considered that all the signal falls in the ON regions since this is already taken into consideration in the effective area curves and we do not consider energy dispersion effects. Moreover, the effective areas and background rates are considered for standard reported analysis configurations, while collaborations can opt for either loose, standard, or hard configurations and can have different cuts for observations taken under moderate moonlight. The lack of public information on telescope effective areas at different zenith angles led us to use scaling factors across different telescope configurations, which can result in further systematic caveats.   

\begin{table*}[!htb]
\centering
\begin{tabular}{ccccccc}
\hline
GRB Name & z & Time  & Obs. delay & Obs. duration & $\mathrm{\sigma}$ $>$ 2000s (full) & $\mathrm{\sigma}$ $>$ 2000s (full) \\
 &  & (UTC) & (s) & (s) & H.E.S.S. \RNum{2} & H.E.S.S. \RNum{1} \\
\hline
GRB060904B & 0.7029 & 2006-09-04T02:31:03 & 26.0 & 5045.0 & $<$1 (1.8) & \textbf{$<$1} (\textbf{$<$1})\\
GRB100621A & 0.542 & 2010-06-21T03:03:32 & 40.0 & 4733.0 & 2.4 (19.6) & \textbf{$<$1} (\textbf{5.7})\\
GRB130925A & 0.348 & 2013-09-25T04:11:24 & 60115.0 & 5303.0 & \textbf{2.4 (2.4)} & 1.0 (1.0) \\
GRB131030A & 1.293 & 2013-10-30T20:56:18 & 27.0 & 8313.0 & \textbf{$<$1} \textbf{(2.0)} & $<$1 ($<$1) \\
GRB161219B & 0.1475 & 2016-12-19T18:48:39 & 388.0 & 11899.0 & \textbf{11.5} \textbf{12.1} & 7.7 (8.0) \\
GRB180720B & 0.654 & 2018-07-20T14:21:44 & 35209.0 & 15302.0 & \textbf{2.5 (2.5)} & $<$1 ($<$1)\\
GRB190829A & 0.0785 & 2019-08-29T19:56:44 & 12179.0 & 16817.0 & \textbf{31.5 (31.5)} & 24.6 (24.6) \\
\hline
\hline
\end{tabular}
\caption{GRBs potentially detectable by the H.E.S.S. telescopes. The name, the redshift and the burst time of the GRB are displayed with the observation delay since the time of burst, the observation duration and the significance of detection by arrays \RNum{2} and \RNum{1}. The significance in bold indicates the array existed at the time of the GRB.}
\label{tab:GRB_HESS}
\end{table*}

\begin{table*}[!htb]
\begin{tabular}{ccccccc}
\hline
GRB Name & z & Time & Obs. delay & Obs. duration & $\mathrm{\sigma}$ $>$ 2000s (full) &  $\mathrm{\sigma}$ $>$ 2000s (full) \\
 &  & (UTC) & (s) & (s) & MAGIC \RNum{2} & MAGIC \RNum{1} \\
 \hline
GRB090417B & 0.345 & 2009-04-17T15:20:03 & 11456.0 & 20150.0 & 2.0 (2.0) & \textbf{1.6 (1.6)} \\
GRB101225A & 0.847 & 2010-12-25T18:37:45 & 1124.0 & 5622.0 & 4.9 (5.4) & \textbf{4.8} (\textbf{5.4}) \\
GRB130427A & 0.3399 & 2013-04-27T07:47:57 & 39056.0 & 2424.0 & \textbf{2.0 (2.0)} & 1.5 (1.5) \\
GRB190829A & 0.0785 & 2019-08-29T19:56:44 & 14906.0 & 11363.0 & \textbf{9.7 (9.7)} & 9.3 (9.3)  \\
\hline
\hline
\end{tabular}
\caption{GRBs potentially detectable by the MAGIC telescopes. Same as Tab.~\ref{tab:GRB_HESS}.}
\label{tab:GRB_MAGIC}
\end{table*}

\begin{table*}[!htb]
\begin{tabular}{ccccccc}
\hline
GRB Name & z & Time & Obs. delay & Obs. duration & $\mathrm{\sigma}$ $>$ 2000s (full) &  $\mathrm{\sigma}$ $>$ 2000s (full) \\
 &  & (UTC) & (s) & (s) & VERITAS \RNum{2} & VERITAS \RNum{1} \\
 \hline
GRB060218 & 0.03342 & 2006-02-18T03:34:30 & 104.0 & 5766.0 & 63.7 (69) & \textbf{51.2} (\textbf{56})  \\
GRB090618 & 0.54 & 2009-06-18T08:28:29 & 31.0 & 8553.0 & 1.5 (1.6) & \textbf{$<$1} (\textbf{$<$1}) \\
GRB190829A & 0.0785 & 2019-08-29T19:56:44 & 46116.0 & 10605.0 & \textbf{6.0} \textbf{(6.0)} & 4.9 (4.9) \\
\hline
\hline
\end{tabular}
\caption{GRBs potentially detectable by the VERITAS telescopes. Same as Tab.~\ref{tab:GRB_HESS}.}
\label{tab:GRB_VERITAS}
\end{table*}

For each instrument, we identify a list of interesting GRBs. 
For H.E.S.S.: GRB 060904B, 080605, 100621A, 100814A, 130925A, 131030A, 161219B, 180720B, 190829A and 210721A. For MAGIC: GRB  060904B, 080605, 090112, 090417B, 101225A, 130430A, 131030A, 190829A and 210619B. For VERITAS: GRB 060218, 090618, 120729A and 190829A. We take a closer look at the light curves of these GRBs. Some of the observation windows for these GRBs cover the prompt and early afterglow phases. During these phases, the VHE gamma-ray light curves might not follow the power law decay presented in sec.~\ref{sec:assumptions}. VHE gamma rays in these early phases have been observed~\citep{LHAASO2023, 2019Natur.575..455M} with photon indices around $\mathrm{\gamma \sim 2}$. Yet, they might show a rise and fall in the emission. Our assumptions are at best valid for the afterglow phase, which is why we introduced a temporal cut at 2000 seconds for the analysis. After 2000 seconds, we fit the power law decay with the remaining X-ray data. In some cases, we verify that the resulting fit does not exceed the X-ray data during the observation window in order not to overestimate the detection significance.  

In Tab.~\ref{tab:GRB_HESS},~\ref{tab:GRB_MAGIC} and~\ref{tab:GRB_VERITAS}, we show the results of the analysis for H.E.S.S., MAGIC, and VERITAS. The significance of detection is shown for both array configurations considered for the three IACT arrays. For each GRB, the significance highlighted in bold is the most realistic, since it would have occurred with the array configuration that was available at the time of the burst. For example, we see that GRB\,100621A, which occurred in 2010, is detectable by H.E.S.S. \RNum{2} at $\mathrm{2.4 \sigma}$ and by H.E.S.S. \RNum{1} at $\mathrm{< 1 \sigma}$. In reality, the H.E.S.S. \RNum{2} configuration did not exist in 2010. For all the results shown here, we use $\mathrm{F= 3}$.  We note that the significance computed is for the full observation window possible after 2000 seconds and might diverge from the significance found for the VHE GRBs by MAGIC and H.E.S.S. Due to the decaying nature of the emission, some collaborations might choose to analyze the first few minutes of observation separately. GRB\,190114C does not appear here as it occurred outside the observation conditions presented in Tab.~\ref{tab:obs_cond}.

\begin{figure*}
  \centering
  \begin{minipage}[b]{0.45\textwidth}
    \includegraphics[width=\textwidth]{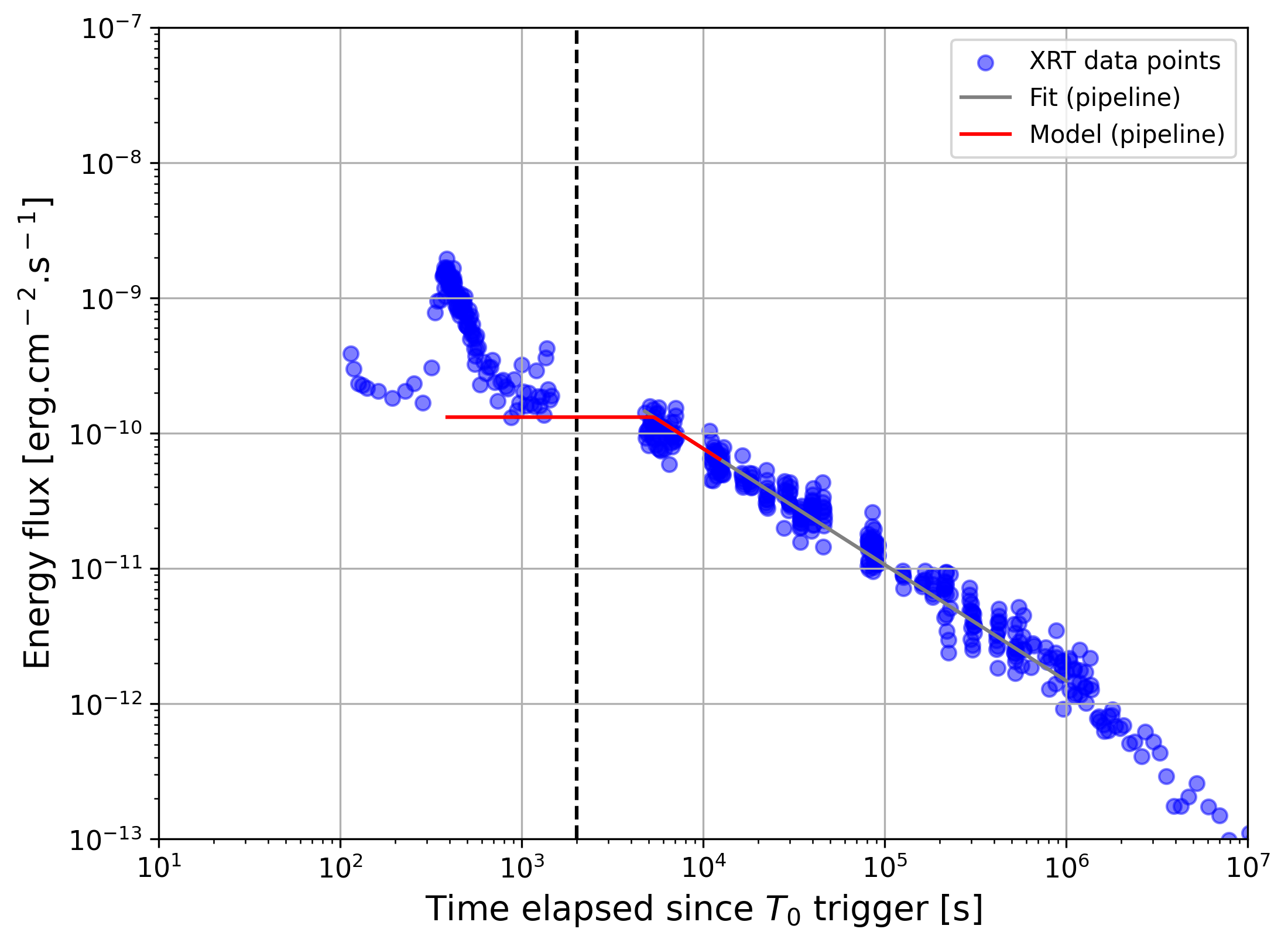}
  \end{minipage}
     \begin{minipage}[b]{0.45\textwidth}
    \includegraphics[width=\textwidth]{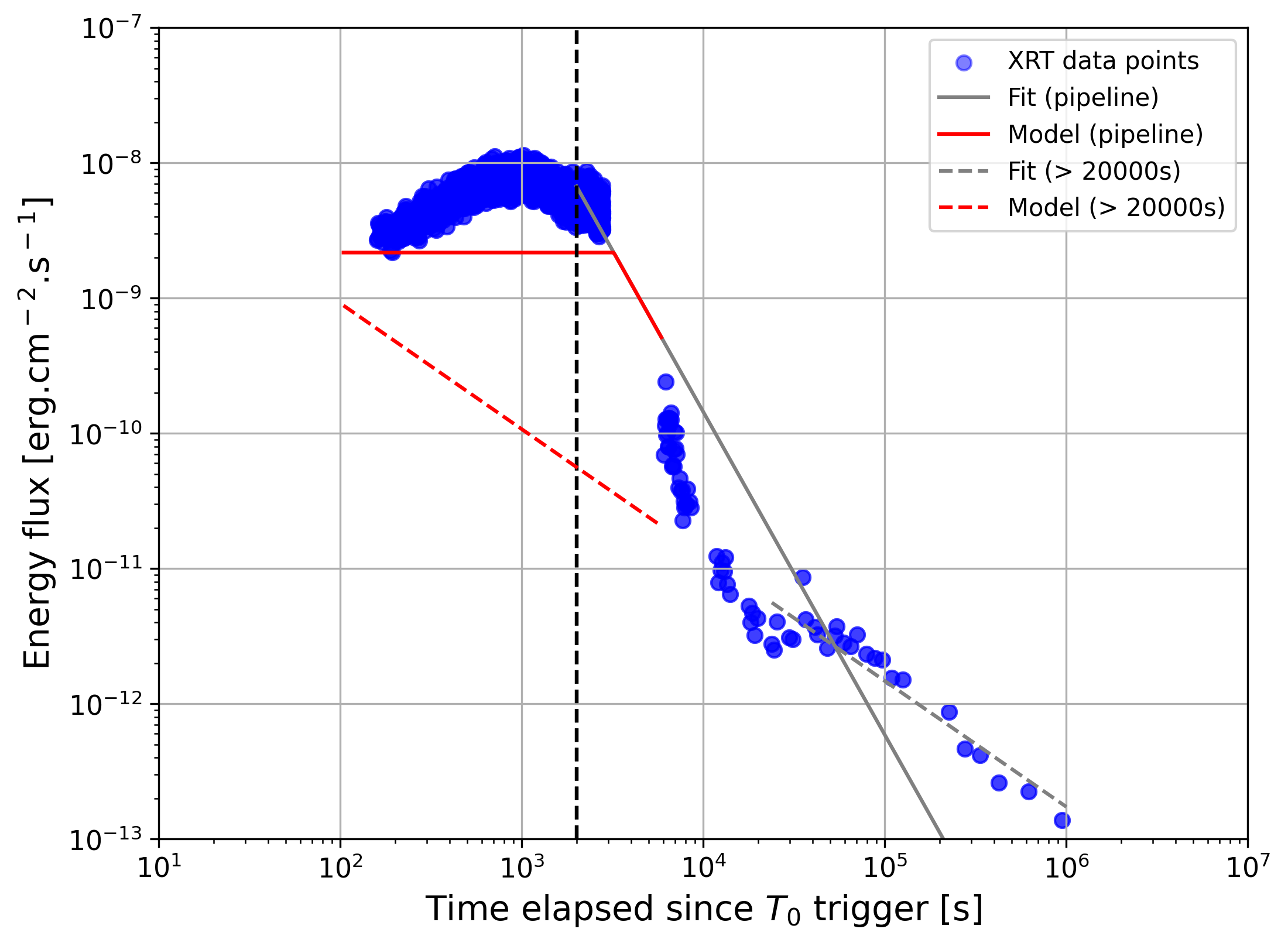}
  \end{minipage}
      \caption{X-ray fits for GRB\,161219B (left) and 060218 (right). The blue dots represent the \emph{Swift}-XRT data. In the pipeline, the minimum between the lowest X-ray point and the fit of the data is used for the computation of the expected VHE signal during the H.E.S.S. (left) and VERITAS (right) observation window represented in red. The dotted gray and red lines on the right represent a fit after 20000 seconds. The dotted vertical black line represents the 2000-second threshold.}
    \label{fig:xray_fit}
\end{figure*}

This work aims at identifying potentially interesting GRBs for IACTs and not to claim exact detection significance. For that reason, we also add in Tab.~\ref{tab:GRB_HESS},~\ref{tab:GRB_MAGIC} and~\ref{tab:GRB_VERITAS} a tentative significance of detection for the potential VHE GRBs that can be observed at early phases, taking into consideration the full observation window. This significance takes into account the data gathered at early times when the power law decay is no longer valid. We ensure that the X-ray emission at early times is always above the fit in the early phase as shown for GRB\,060218, and 161219B in Fig.~\ref{fig:xray_fit}. In light of the aforementioned, the significance reported for the whole observation window can be considered as a lower limit.

We find that GRB\,060904B and 131030A might have been detectable by current generation IACTs provided only that the observation started early on, while GRB\,060218, 090618, 090417B, 100621A, 101225A, 130427A, 130925A,  161219B, 180720B, and 190829A could also have been detected at later times by any of the historical 6 array configurations. GRB\,060904B, and 090618 would not have been detected by the H.E.S.S. and VERITAS arrays that were online at the time, respectively. We highlight GRB\,161219B, which shows significant signs of VHE gamma rays in the afterglow phase, and highly recommend IACT collaborations to look at any data available on this GRB. GRB\,101225A, and 060218 also show potentially highly significant detections at later times. We find the highest significance of detection for GRB\,060218 at $\mathrm{> 50 \sigma}$ for all cases. We note that the VERITAS science operations with the full array started in 2007, after the detection of this GRB. Fig.~\ref{fig:xray_fit} shows that most of the observation window falls prior to the steady decay afterglow phase for which our initial assumptions were made. From the X-ray data, we estimate the steady decay to start at 20000 seconds. We refit the X-ray data after this time and find a significance of $\mathrm{2.9 \sigma}$  and $\mathrm{7.7 \sigma}$ for the full window and the window after 2000 seconds respectively. 

From the literature, integral upper limits of $\mathrm{4.2 \times 10^{-12} cm^{-2} s^{-1}}$ above 380 GeV were reported by H.E.S.S. on GRB\,100621A~\citep{hess2014}. Observations started only 700 seconds after the trigger and lasted one hour. In the same study, the gamma-ray flux is constrained to be at least 2.5 times lower than the X-ray flux. GRB\,130427A was known as the brightest of all times before 2022, and is one of the longest GRBs (20 hours). VERITAS took a 16485-second exposure with a $\sim$20 h delay. No significant detection was found but integral upper limits of $\mathrm{3.3 \times 10^{-12} cm^{-2} s^{-1}}$ above 100 GeV were derived~\citep{Aliu_2014}. A record-breaking 95 GeV photon was recorded by Fermi-LAT a few minutes after the burst~\citep{Ackermann_2014}. Although many researchers believed that GRBs emit TeV gamma-rays, it was still an open question until the first discovery was confirmed in 2019. GRB\,060218 is reported by~\cite{He_2009} to be a low luminosity burst with an expected high gamma-ray flux at early times followed by a rapid decay of the emission (as seen from the X-rays). No reported observations were found on GRB\,060218, 060904B, 090417B, and 090618 by H.E.S.S., MAGIC or VERITAS prior to 2012~\citep{Aharonian_2009,2009arXiv0907.1001G,Acciari_2011}.

\section{Discussion}
\label{sec:discussion}
The results presented in Sec.~\ref{sec:results} show that, given the assumed relationship between the VHE and the X-ray fluxes, some VHE GRBs may have gone undetected by IACTs since 2004. We highlighted in Sec. \ref{sec:results} only the GRBs with known redshift, acknowledging that most GRBs do not have redshift measurements. Therefore, the number of detectable VHE GRBs likely exceeds that determined in our study. On the other hand, it is the brightest GRBs that benefit from redshift measurements in real life. We note the importance of the upgrades to IACTs, which improve their effective areas at low energies and hence, their ability to detect GRBs. The upgrade from H.E.S.S. \RNum{1} to H.E.S.S. \RNum{2} allows to lower the energy threshold from 200 GeV to 100 GeV. For example, after 2012, GRB\,130925A, 131030A, and 180720B  would not be as significantly detectable with the H.E.S.S. \RNum{1} array at a 200 GeV energy threshold. We see also that some GRBs reported in Sec. \ref{sec:results} are flagged as interesting for more than one IACT and more than one configuration. The question remains, why no VHE gamma rays were ever reported from these GRBs? To answer this question, we highly encourage the three IACT collaborations to look for any data that might have been collected on the reported GRBs. Even if no significant VHE gamma rays are observed, the study of these GRBs can constrain the relation between the X-ray and the VHE gamma-ray fluxes. On this note, we highlight the importance of publishing and updating telescope performance data (such as effective areas and background rates) for studies comparing theoretical and experimental results. 

\begin{figure*}[!ht]
  \centering
\includegraphics[width=0.8 \textwidth]{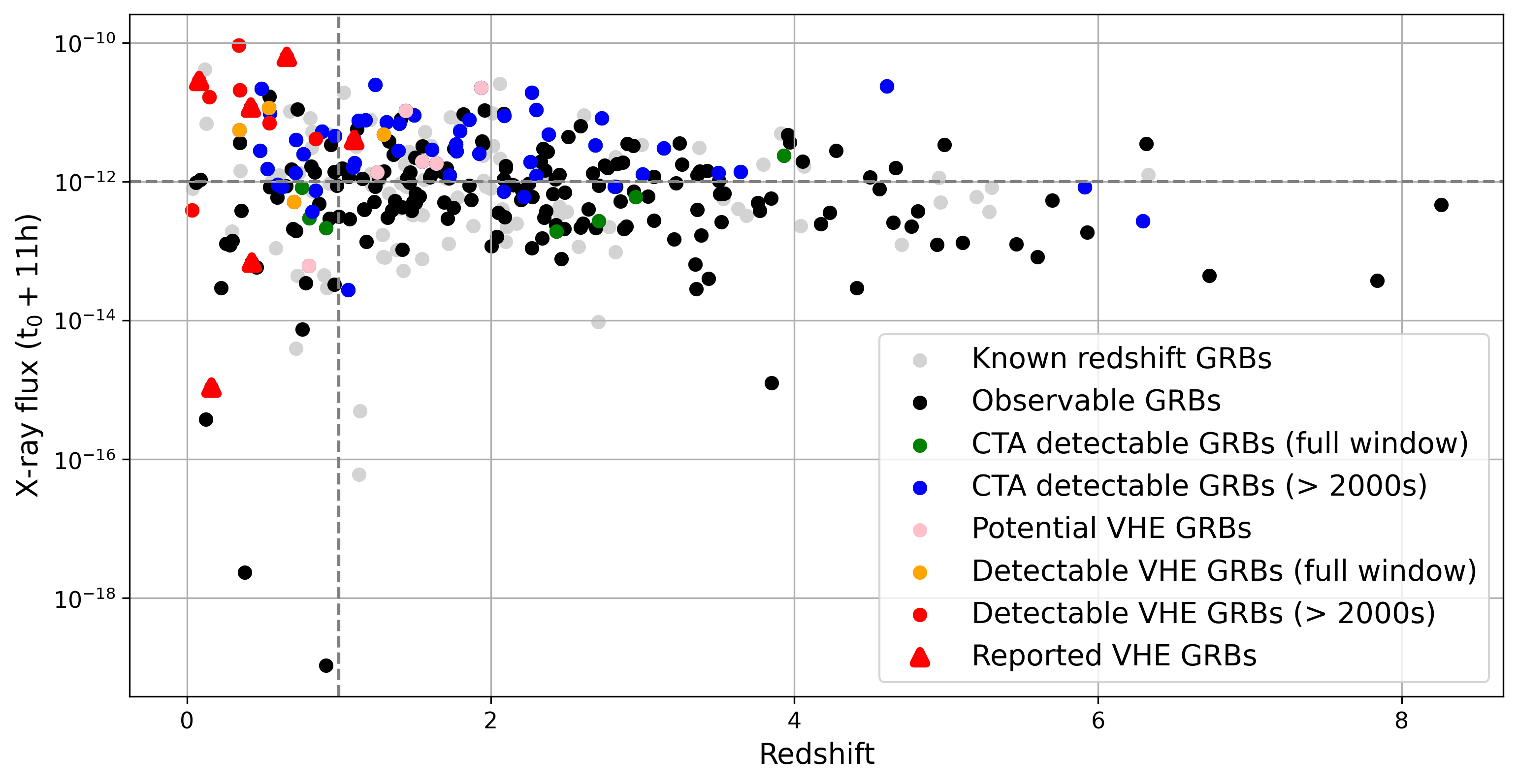}
\caption{GRB X-ray fluxes at 11 hours vs. redshift. The gray dots represent all GRBs that have a redshift measurement and an XRT X-ray flux at 11 hours (348 GRBs in total). The black dots represent the ones amongst them that are observable by the H.E.S.S., MAGIC, or VERITAS IACTS (248). The red triangles represent the GRBs for which VHE gamma-ray detections have been reported (6). The red dots represent in addition the GRBs that are detectable by current IACTs in the afterglow phase using data after 2000 seconds (8). The orange dots represent in addition the GRBs for which a prompt observation is possible and is detectable if data from the entire observation window is used (12). The pink dots represent in addition the GRBs that were initially flagged by our study as being potentially detectable at VHE energies (18). The blue and green dots are for CTA the same as the orange and red dots respectively (65 and 71).}
\label{fig:XZplot}
\end{figure*}

As expected, low redshift GRBs are favored for VHE gamma-ray detection, since their spectrum is less affected by EBL absorption. High energy fluxes are naturally also favored. The plot in Fig.~\ref{fig:XZplot} (xz plot in the following) shows the redshift versus the X-ray flux at 11 hours taken from the XRT GRB catalog~\citep{2009MNRAS.397.1177E} (when available). We show in light gray all \emph{Swift} GRBs that have a redshift measurement and in dark gray the ones that are observable by at least one of the three IACTs. The first quadrant featuring high X-ray emission and low redshift GRBs is populated by many GRBs with the highest significance of detection in the afterglow phase (red dots). The GRBs detected in real life by IACTs can also be found mostly in this quadrant (red triangles). The GRBs that are potentially detectable with early observations only (orange dots) and the GRBs flagged as interesting in this study (pink dots) extend to the second and fourth quadrants having either high X-ray emission or low redshift. The second quadrant features GRBs that can be detected by IACT configurations with low energy thresholds.

Although some GRBs might potentially have been detected at VHE in the past, their number is low with current IACT capacities. The rate of interesting GRBs for the VHE community with current IACT capabilities can be concluded to be less than 1 per year (between 0.6 and 0.8 per year). This includes the GRBs shown in Sec.~\ref{sec:results} and the three other GRBs for which MAGIC reported detection or a candidate detection: GRB\,190114C, 201015A~\citep{2020GCN.28659....1B}, 201216C~\citep{2024MNRAS.527.5856A, 2020GCN.29075....1B, 2023ApJ...952..127Z} and 160821B~\citep{2021ApJ...908...90A}. IACT observation and visibility conditions and the availability of X-ray and redshift measurements are taken into account in this number. Therefore, the overall number of VHE GRBs might be higher. Moreover, non-ideal observation conditions, like observations at high zenith angles, short observation periods, and long observation delays, affect this number. With such a low rate of VHE GRBs, a low IACT duty cycle, and a need for ideal observation conditions (darkness, low zenith angle), it is not surprising that no significant detections were reported for many years. While low IACT duty cycles cause the low rate of GRB detections at VHE, we see here that additional possible reasons might in reality play a significant role, including weather conditions and strict follow-up observation criteria. Amongst these factors, the maximal allowed observation delay may be the one that has proved most in need of further loosening, especially considering GRB\,180720B and GRB\,190829A, detected hours and days after the burst. While short observation delays are shown to be clearly favored for a VHE detection (highlighting the importance of fast reaction and slewing time), in this study, GRB\,101225A, 130427A, 130925A, 161219B, 180720B, and 190829A might be detectable with several hours of observational delays.

The question of why five detection were reported in a span of 2 years while none were reported before can be due to the fact that IACT collaborations loosened their observation criteria in recent years. For example, GRB\,180720B was only observed $\sim$10 hours after the burst. Moreover, GRB\,19011AC observations were pushed to the limit with the MAGIC collaboration, which chose to observe it during moonlight and at high zenith angles. These observations highlight the endeavors towards the capability to measure the prompt phase. After 2020, the detection rate went down to zero again for two years. Yet, during these two years, the brightest two GRBs of all times, GRB\,221009A and 230307A were detected by X-ray instruments ~\citep{2023ApJ...946L..31B, 2023ApJ...946L..24W, 2023GCN.33411....1D, 2023GCN.33414....1B}. However, they unfortunately fell outside IACT observation conditions. GRB\,221009A at $\mathrm{z = 0.151}$~\citep{2022GCN.32686....1C, 2022GCN.32648....1D, 2022GCN.32765....1I} is the brightest GRB of all times and it occurred during the full moon, preventing IACTs from acquiring data until a few days later with no significant detection~\citep{2023ApJ...946L..27A}. This suggests that the relation between the X-rays and the gamma rays might not hold for late times and for all cases.

We repeat our study for the Cherenkov Telescope Array (CTA) observatory, considering CTA to have been functional since 2004. This exercise represents a purely hypothetical scenario for exploratory purposes. CTA is the future IACT planned to be an order of magnitude more sensitive than current generation telescopes and a lower energy threshold. It will be built on two sites: a northern site in La Palma and a southern site in Chile. The southern site is estimated to have an energy threshold as low as 60 GeV. Four Large Sized Telescopes (LSTs) will be installed in the northern site lowering the energy threshold to 20 GeV. We use the effective area and background rates found on the CTA observatory website\footnote{\url{https://www.cta-observatory.org/science/ctao-performance/}}~\citep{2013APh....43..171B}. We do not aim to establish prospects for CTA. We only show in a qualitative manner what CTA could have done in comparison with current IACTs. Therefore, we consider the same effective area and background rate for all zenith-angle observations. In addition to the VHE GRBs found in Fig.~\ref{fig:XZplot}, CTA has the ability to extend the detection possibilities far into the second quadrant in the xz plot. This is mainly due to its capability to detect low-energy VHE gamma rays, which are less affected by EBL absorption. CTA's higher sensitivity minorly impacts the detection capabilities in the fourth and third quadrants. However, the overall detection rate with the two sites increases by half an order of magnitude. These numbers should be considered carefully, since there are three IACTs on three different sites while CTA will be on two sites, lowering the number of GRBs falling within observation and visibility criteria. We note also that the rate of redshift measurement in the northern and southern hemispheres is not homogeneous. An enhancement to this study  would be to compare the capabilities of CTA's northern site to MAGIC, as they both share the same site.  We note that a GRB is more probable to happen at a large zenith angle, given the larger solid angle, resulting in higher telescope energy thresholds. As only one observation configuration is considered for CTA, these rough, non-conservative numbers are only indicative. 

\section{Summary}
\label{sec:conclusion}
In this work, we study the case of the missing GRBs in the VHE domain. To do so, we consider three IACTs (H.E.S.S., MAGIC, and VERITAS) to identify for each of them independently which GRBs they could have observed in the VHE domain since 2004 :
\begin{itemize}
\item	12 GRBs are identified to be most likely to be detected in the VHE domain by H.E.S.S., MAGIC, or VERITAS in the past 18 years.
\item As expected, GRBs with low redshift and high X-ray emission are favored for detection, as well as low observation delays.
\item Accounting for low IACT duty cycle and performance at low energies, the rate of GRBs detectable at VHE with current IACT capabilities is low and is estimated to be $\mathrm{<1}$ per year. Weather conditions and hardware malfunctions are not accounted for. The time of the installation of automatic telescope reaction to GRB alerts is also not considered and could significantly influence the statistics of potential GRB detections.
\item Loosening observation criteria and automatic telescope reaction are crucial for VHE GRB detection. The coverage of several time zones in the two hemispheres to increase coverage and reduce observation latency is recommended.   
\item	We repeat the study for the two sites of CTA and GRBs detected between 2004 and June 2022. We find that the number of potentially detectable GRBs increases significantly with only two sites. 
\end{itemize}

We independently establish that bright emission, low redshift, short delays, and deep observations are in favor of detecting a GRB in the VHE domain. The low IACT duty cycle is a major reason for the past years' lack of VHE GRB detections. Loosening follow-up observations is crucial to increase detections. With a lower energy threshold, CTA is able to increase the detection rate considerably.  The relation between the X-rays and the VHE gamma rays is deduced from a few observations of GRBs at VHEs. However, additional data might show significant fluctuations in the established relations. H.E.S.S. observations of GRB\,2201009A at 51 hours after the burst reported no significant detections, suggesting that the relation might not hold at this timescale. This work should trigger deepened investigations of any data on retrospectively detectable VHE GRBs, allowing to better constrain GRB afterglow physics.

\section*{Acknowledgments}
This work made use of data supplied by the UK Swift Science Data Centre at the University of Leicester and of the H.E.S.S. ToO Alert system software. We thank the H.E.S.S. collaboration and the LLR Astrogamma group members for fruitful discussions of this work. 

\bibliographystyle{aasjournal}
\bibliography{main}

\end{document}